\documentclass[pdflatex,sn-mathphys-num]{sn-jnl}

\usepackage{graphicx}%
\usepackage{multirow}%
\usepackage{amsmath,amssymb,amsfonts}%
\usepackage{amsthm}%
\usepackage{mathrsfs}%
\usepackage[title]{appendix}%
\usepackage{xcolor}%
\usepackage{textcomp}%
\usepackage{manyfoot}%
\usepackage{booktabs}%
\usepackage{algorithm}%
\usepackage{algorithmicx}%
\usepackage{algpseudocode}%
\usepackage{listings}%
\usepackage{qcircuit}
\usepackage{subfig}
\usepackage{booktabs} 
\def\be{\begin{equation}}
\def\ee{\end{equation}}
\def\bea{\begin{eqnarray}}
\def\eea{\end{eqnarray}}
\usepackage{braket}
\usepackage{soul}
\usepackage{comment}

\def\2{\frac{1}{2}}

\def\4{\frac{1}{4}}

\theoremstyle{thmstyleone}%
\theoremstyle{thmstyletwo}%

\theoremstyle{thmstylethree}%

\begin{document}

\title[Article Title]{Continuous-variable Quantum Boltzmann Machine}


\author*[1]{\fnm{Shikha} \sur{Bangar}}\email{sbangar@vols.utk.edu}

\author[1]{\fnm{Leanto} \sur{Sunny}}\email{lsunny@vols.utk.edu}

\author[2]{\fnm{K\"ubra} \sur{Yeter-Aydeniz}}\email{kyeteraydeniz@mitre.org}

\author[1]{\fnm{George} \sur{Siopsis}}\email{siopsis@tennessee.edu}

\affil*[1]{\orgdiv{Department of Physics and Astronomy}, \orgname{University of Tennessee}, \orgaddress{\street{1408 Circle Drive}, \city{Knoxville}, \state{Tennessee}, \postcode{37996-1200},  \country{USA}}}

\affil[2]{\orgdiv{Emerging Technologies and Physical Sciences}, \orgname{MITRE Corporation}, \orgaddress{\street{7515 Colshire Drive}, \city{McLean}, \state{Virginia}, \postcode{22102-7539}, \country{USA}}}



\abstract{
We propose a continuous-variable quantum Boltzmann machine (CVQBM) using a powerful energy-based neural network. It can be realized experimentally on a continuous-variable (CV) photonic quantum computer. We used a CV quantum imaginary time evolution (QITE) algorithm to prepare the essential thermal state and then designed the CVQBM to proficiently generate continuous probability distributions. We applied our method to both classical and quantum data. Using real-world classical data, such as synthetic aperture radar (SAR) images, we generated probability distributions. For quantum data, we used the output of CV quantum circuits. We obtained high fidelity and low Kuller-Leibler (KL) divergence showing that our CVQBM learns distributions from given data well and generates data sampling from that distribution efficiently. We also discussed the experimental feasibility of our proposed CVQBM. Our method can be applied to a wide range of real-world problems by choosing an appropriate target distribution (corresponding to, e.g., SAR images, medical images, and risk management in finance). Moreover, our CVQBM is versatile and could be programmed to perform tasks beyond generation, such as anomaly detection.}




\maketitle

\section{Introduction}\label{section:intro}
Boltzmann machines (BMs) are robust machine learning models which provide modeling of probability distributions. BMs consist of a probabilistic network that is represented as an undirected graph with hidden and visible nodes. In this machine learning model, the probability distribution of the data is approximated based on a finite set of samples. The information is encoded into the bias coefficients and weights of the neural network. After a successful training process, the learned distribution resembles the actual distribution of the data  that can make correct predictions about unseen instances. However, the generalization of the model suffers from a growing number of parameters and the training of a classical BM can become impractical. For this reason, it is worth exploring engineering quantum Hamiltonian Ans\"atze to construct quantum BMs \cite{amin2018quantum}. In QBMs, the goal is to approximate the thermal state of the Hamiltonian Ansatz to the target density matrix of the distribution through training of the weights of the Hamiltonian Ansatz. One of the advantages of QBMs over their classical counterparts is that QBMs have higher expressive power, i.e., can fit a wider range of complex data patterns, since the Hamiltonian can include more non-commuting terms \cite{coopmans2023sample}. 

Preparing a thermal state and computing its properties is a challenging task. Hence, various heuristic methods that are compatible with noisy intermediate scale quantum (NISQ) hardware have been proposed \cite{huijgen2023training}. These methods include quantum imaginary time evolution (QITE) \cite{motta2020determining}, variational QITE (varQITE) \cite{McArdle_2019}, and the variational quantum eigensolver  ($\beta-$VQE) \cite{liu2021solving}. The thermal states prepared utilizing these tools have also been used in realizing QBMs. For instance, in Ref.\ \cite{zoufal2021variational}, a variational QBM was proposed and utilized for a discriminative learning task, specifically fraud detection, as well as a generative learning task, specifically learning a Bell state probability distribution. The method was implemented on IBM Quantum hardware. This work was followed by several studies including \cite{huijgen2023training,Shingu_2021}. In Ref.\ \cite{Shingu_2021}, similar to \cite{zoufal2021variational}, they utilized variational QITE evolution to study QBMs and tested their model using the bars and stripes (BAS) dataset. They focused on the classical cases when the Hamiltonian for the QBM has only diagonal terms in the computational basis, and reduced the required number of qubits by half by preparing an initial pure state which was a superposition of all states. On the other hand, in Ref.\ \cite{huijgen2023training}, $\beta-$VQE was used as a subroutine for preparing the Gibbs state needed for QBM. In their QBM model, free energy, or equivalently quantum relative entropy, was used as the loss function. They tested their model both on quantum and classical data. It was proposed that, once the classical dataset is embedded into the quantum circuit, a lower rank $\beta-$VQE can be utilized for training of QBM. A similar feature was also observed for low temperature quantum data. However, for higher temperature quantum data, a higher rank $\beta-$VQE is required. They also tested their QBM model on a real-world dataset, namely salamander retina, and demonstrated that $\beta-$VQE QBM achieves lower Kuller-Leibler (KL) divergence value compared to a classical Boltzmann machine. In Ref.\ \cite{coopmans2023sample}, the sample complexity of QBM machine learning was studied. It was shown that QBMs do not possess barren plateaus (optimization landscape becoming flat due to various reasons \cite{mcclean2018barren}) and that pre-training on a subset of the QBM parameters can only lower the sample complexity bounds. In addition to gate based quantum computing, QBMs have been demonstrated using quantum annealers. In Ref.\ \cite{dixit2021training}, a restricted QBM was utilized for balancing the imbalanced credit card transactions data through synthetic data generation that can be utilized for fraud detection and other cyber security applications.

Studies of quantum generative models have mainly focused on the generation of discrete probability distribution instances. One example is Ref.\ \cite{romero2021variational}, where a variational quantum generative adversarial network (GAN) model was utilized for the generation of the probability distribution of a real dataset obtained using a variational quantum generator with fixed parameters. In Ref.\ \cite{anand2021noise}, this method was shown to be robust to quantum hardware noise by implementing it on Rigetti quantum hardware. However, most real-world applications, such as image and sound generation, computational fluid dynamics, study continuous probability distributions. Use of discrete-variable (DV) quantum computing (based on qubits) is only partially suited for continuous-valued data. One can avoid the burden of encoding continuous valued classical data into DV by using CV quantum machine learning tools. For instance, in Ref.\ \cite{anand2024time}, they studied time series forecasting for continuous datasets, such as energy consumption and stock price data.   It is more natural to extend generative models to continuous-variable quantum computing (CVQC) models for the study of continuous data. In CVQC, quantum information is encoded into infinite-dimensional qumodes, such as the electromagnetic field of photons in photonic quantum hardware. In Ref.\ \cite{vcepaite2022continuous}, a CV quantum Born machine was developed which offered an efficient way of modeling continuous probability distributions, such as a Gaussian distribution. By choosing a Gaussian kernel, it was demonstrated that a CV quantum Born machine naturally provides a much better fit to a Gaussian distribution with a single qumode compared to a 10-qubit quantum circuit Born machine model. It was also shown that a 3-qumode Gaussian and cubic kernel CV quantum Born machine learns a quantum target distribution much faster than by using a classical Gaussian kernel method.


In this work, we discuss a proposal for a continuous-variable quantum Boltzmann machine (CVQBM). Our CV model is a natural fit for studying continuous probability distributions. We utilize our method for generative modeling in real-world applications, such as SAR image probability distribution generation, as well as probability distribution of data obtained from prepared quantum states. To the best of our knowledge, this is the first demonstration of generation of various real-world applications' continuous probability distributions using a CV quantum generative model. Even though experiments were conducted on a simulator, we discuss experimental feasibility of our simulations and demonstrate that problems studied in this work can be implemented on experimentally feasible photonic quantum computing architectures. In our model, the CV quantum circuit we employ demonstrates a CV quantum generative model, i.e., CVQBM, and also presents a variational CV QITE algorithm utilized for finite temperature state preparation. Our variational CV QITE algorithm can be leveraged in other use cases, such as ground state preparation in quantum field theories \cite{yeter2022quantum}.  

Our discussion is organized as follows. In Section \ref{section:methods}, we provide the details of the proposed CVQBM model. We design the CVQBM quantum circuit and discuss the procedure for setting it up as a continuous probability distribution generator. In Section \ref{sec:classicaldata}, we present various case studies using our CVQBM, focusing on classical data from the SAR images. We generate data for forest and sea monitoring. In Section \ref{sec:quantumdata}, we discuss quantum data, including both Gaussian and non-Gaussian cases. An essential aspect of our proposed CVQBM is that it is experimentally feasible on CV photonic quantum computers, which is discussed in Section \ref{sec:exp}. We introduce noise to analyze the robustness of our CVQBM, and discuss the success rate and gate parameters chosen in our case studies. Finally, in Section \ref{sec:conclude}, we present our conclusions and indicate future directions.

\section{The Method} \label{section:methods}

In this Section, we outline the design of a quantum circuit for the CVQBM and set up the machine as a continuous distribution generator. We show that the output can be thought of as a thermal state of an effective Hamiltonian defined by the quantum circuit. This is similar to 
the approximate Gibbs state that was prepared using a variational quantum imaginary time evolution (varQITE) algorithm with discrete variables \cite{zoufal2021variational}. 

The CVQBM circuit is designed to prepare a thermal state, $\frac{1}{\mathcal{Z}} e^{-\beta H(\Vec{\zeta})}$, where $\mathcal{Z} = \mathrm{Tr} e^{-\beta H(\Vec{\zeta})}$, for a Hamiltonian that depends on parameters collectively denoted as the vector $\vec{\zeta}$. For simplicity, we set the inverse temperature $\beta =1$. This state represents quantum imaginary time evolution (QITE) which is not unitary. The parameters $\vec{\zeta}$ control the quantum gates of the CVQBM circuit, which are unitary, and are trained according to the problem at hand. We achieve non-unitary evolution by performing measurements on ancilla qumodes. Here, we concentrate on a Hamiltonian that depends on a single (visible) qumode of quadratures $(q_v,p_v)$. Extension of our method to include multiple qumodes is straightforward.

\begin{figure}[ht!]
    \centering
    \includegraphics[trim={0.3cm 2cm 6cm 1cm},clip,width=1. \textwidth]{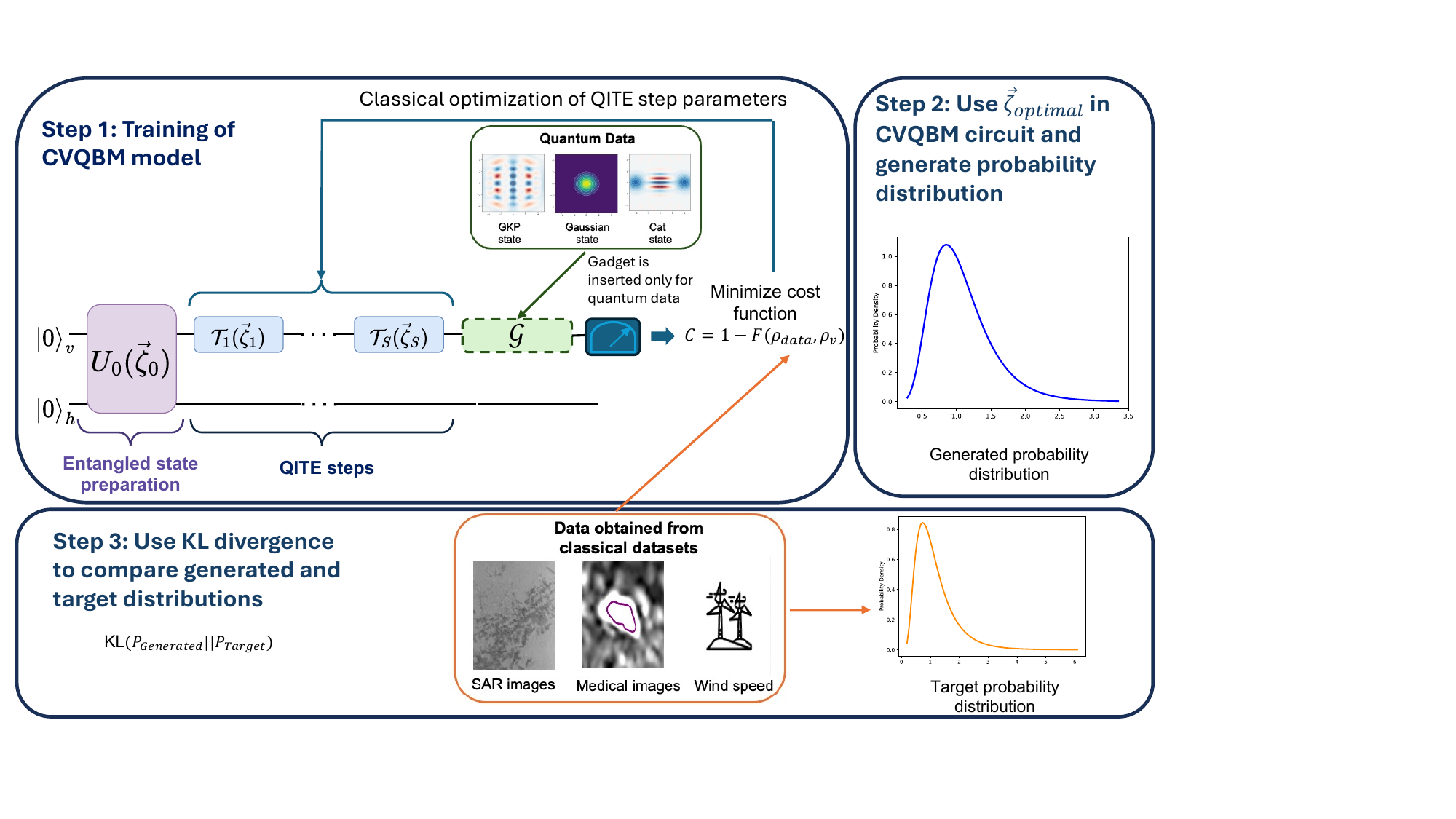}
    \caption{CVQBM framework: Step 1 is the training of the CVQBM model by training the parameters of the quantum circuit, which prepares an entangled state between the visible and hidden modes of the CVQBM, and then performs imaginary time evolution, minimizing the cost function \eqref{eq:cost}, calculated using the density matrix of the data and the generated density matrix. In Step 2, the optimal quantum circuit parameters are used in the same quantum circuit for CVQBM to generate a probability distribution. Step 3 involves verification of how close the generated probability distribution is to the target distribution by using metric KL divergence. For embedding quantum data we also propose a gadget $\mathcal{G}$ (Figure \ref{fig:qgadget}) that is implemented on the visible node following the QITE steps.  }
    \label{fig:CVQBMframework}
\end{figure}

\begin{figure}[!tbp]
  \centering
  \subfloat[]{\includegraphics[width=0.28 \textwidth]{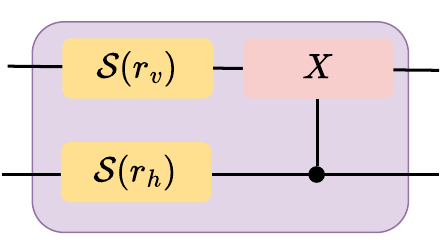}\label{fig:U0det}}
  \hfill
  \subfloat[]{\includegraphics[trim={0cm 29.5cm 35cm 0cm}, clip, width=0.7 \textwidth]{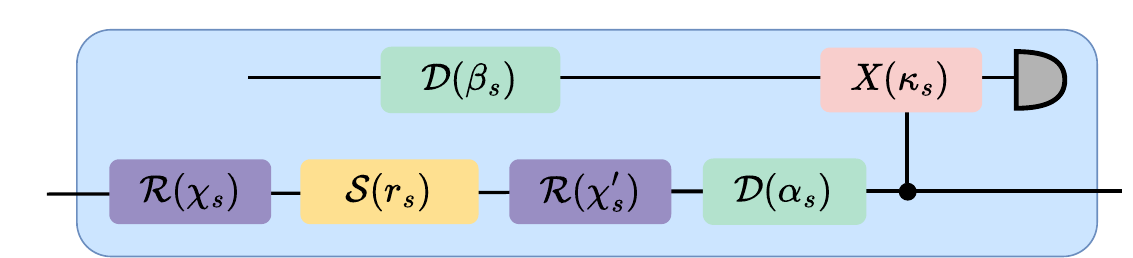}\label{fig:Udet}}
  \caption{Designing the CVQBM circuit, detailed structure of \textbf{(a)}
 $U_0$ where two modes are entangled by using squeezed gates and then applying the CX gate and \textbf{(b)} $\mathcal{T}_s(\vec{\zeta}_s)$ where the gates have trainable parameters collectively represented by $\zeta$. An ancillary mode is added, which introduces non-linearity into the network by entanglement (CX gate) and photon number measurement.} \label{fig:QBMarch}
\end{figure}

The CVQBM circuit is depicted in Figure \ref{fig:CVQBMframework}. In addition to the visible qumode, it uses an ancilla qumode of quadratures $(q_h,p_h)$ representing a hidden mode. Both modes are initially in the vacuum state. They are entangled with the two-mode unitary $U_0$. No further action is taken on the hidden mode. The visible mode is taken through a series of non-unitary quantum gates $\mathcal{T}_s (\Vec{\zeta}_s)$, where $s=1,2,\dots$, that depend on parameters $\Vec{\zeta}_s$ to be trained. Next, we proceed to describe the various quantum gates used in detail.

We start by engineering the entangled state:
\be\label{eq:32} \ket{\psi_0}_{vh} = \frac{1}{\sqrt{\mathcal{N}}} \sum_{n=0}^\infty e^{-\delta n} \ket{n}_v \ket{n}_h \ , \ \ \mathcal{N} = \frac{1}{1 - e^{-2\delta}}\ee
where $\delta >0$ is a fixed constant, and $\ket{n}$ is the eigenstate of the number operator $N$ with $n$ photons. How this is implemented is shown in Figure \ref{fig:U0det}. We use two squeezed modes, one squeezed in the $q$ quadrature with squeezing parameter $r_v = -\frac{1}{2} \log\delta$, and the other one squeezed in the $p$ quadrature with $r_h = \frac{1}{2} \log\delta$. We 
 entangle them with the CX gate $e^{iq_vp_h}$ to bring them in the state
\be \frac{1}{\sqrt{\pi }}\int dq_v dq_h e^{-\frac{\delta}{2} q_v^2} e^{-\frac{1}{2\delta} q_h^2} \ket{q_v}_v \ket{q_v+q_h}_h~. \ee 
For small $\delta$, this is approximately $\ket{\psi_0}$ (Eq.\ \eqref{eq:32}).
This is easily seen by switching to the number operator basis, in which this state can be written as
\be \frac{1}{\sqrt{\pi }} \sum_{n_v,n_h} \int dq_v dq_h e^{-\frac{\delta}{2} q_v^2} e^{-\frac{1}{2\delta} q_h^2} \braket{q_v\vert n_v} \braket{n_h \vert q_v+q_h} \ket{n_v}_v \ket{n_h}_h~,  \ee
expressing the inner products (harmonic oscillator wavefunctions in the position representation) 
as
\be \label{eq:psi} \Psi_n (q) = \braket{q \vert n} = \frac{1}{\sqrt{2^n n!\sqrt{\pi}}} H_n(q) e^{-\frac{1}{2} q^2} \ee 
where $H_n$ is the Hermite polynomial:
\be H_n (q) = (-)^n e^{q^2} \frac{d^n}{d q^n} e^{-q^2} \ee 
and performing the integrals over the $q$ quadratures.


Next, we evolve the state $\ket{\psi_0}$ by acting with a string of non-unitary gates $\mathcal{T}_s (\Vec{\zeta}_s)$ ($s=1,2,\dots, S$) acting on the visible mode only. The total number of gates $S$ depends on the problem at hand. The quantum circuit implementing a gate $\mathcal{T}_s (\Vec{\zeta}_s)$ is depicted in Figure \ref{fig:Udet}. It consists of Gaussian gates only that act on the visible qumode and an ancilla qumode of quadratures $(q_A,p_A)$. Non-unitarity is introduced by a measurement of the ancilla qumode, projecting it onto the state $\ket{n_A=1}_A$. Let $\ket{\psi_{s-1}}_{vh}$ be the input at the $s$th step and denote by $\mathcal{U} (\Vec{\zeta}_s)$ the two-mode unitary representing the collective action of the Gaussian gates on the visible and ancilla qumodes,
\be \mathcal{U} (\Vec{\zeta}_s) = \mathcal{R}(\chi_s) \mathcal{S}(r_s) \mathcal{R}(\chi'_s) \mathcal{D}(\alpha_s) X(\kappa_s) \ee 
where $\mathcal{R}(\chi_s)$ is phase space rotation operator, $\mathcal{S}(r_s)$ represents squeeze gate affecting position and momentum, $\mathcal{D}(\alpha_s)$, is displacement operator and $X(\kappa_s)$ is controlled-X gate, a controlled displacement in position. These gates are defined by the following expressions: 
\begin{equation}
    \mathcal{R} (\chi)  =  e^{ i \chi \hat{a}^{\dagger} \hat{a} }~, \ \ 
    \mathcal{S} (r) =  e^{ \frac{r}{2} \left(  \hat{a}^2 -  \hat{a}^{{\dagger}^2} \right) }, \ \
    \mathcal{D}(\alpha) =  e^{  \alpha (\hat{a}^\dagger -  \hat{a} )} , \ \ 
    X (\kappa)  =  e^{- i \kappa \hat{x} \otimes \hat{p} }~,
\end{equation}
where $\chi, r, \alpha, \kappa \in \mathbb{R}$, $\hat{x} = \frac{1}{\sqrt{2}}(\hat{a}+\hat{a}^\dagger)$, $\hat{p} = -i \frac{1}{\sqrt{2}}(\hat{a}-\hat{a}^\dagger)$, and $\hat{a}$, $\hat{a}^\dagger$ are annihilation and creation operators, respectively. Then the output of the $s$th step is
\be \ket{\psi_s}_{vh} \propto \mathcal{T}_s (\Vec{\zeta}_s) \ket{\psi_{s-1}}_{vh} \ , \ \  \mathcal{T}_s (\Vec{\zeta}_s) = {}_A\braket{1 \vert \mathcal{U} (\Vec{\zeta}_s) \vert 0}_{A}~. \ee
The success probability of obtaining the desired output is $P_s = \| \mathcal{T}_s (\Vec{\zeta}_s) \ket{\psi_{s-1}}_{vh} \|^2$. The circuit must be designed so as to maximize the probability of success.

It should be pointed out that the introduction of an ancilla qumode is similar to the training of CV Quantum Neural Networks (QNN) that are experimentally realizable, as we discussed in our previous work \cite{bangar2023experimentally}. Here we aim at a similar construction that would yield experimentally realizable CVQBM.

After $S$ steps, we obtain the output of the CVQBM,
\be \ket{\mathrm{out}}_{vh} = \ket{\psi_S} \propto \mathcal{T} (\Vec{\zeta}) \ket{\psi_0}_{vh} \ , \ \ \mathcal{T} (\Vec{\zeta}) \equiv \mathcal{T}_S (\Vec{\zeta}_S) \cdots \mathcal{T}_1 (\Vec{\zeta}_1) \ee
where $\Vec{\zeta} \equiv \{ \Vec{\zeta}_1 , \dots , \Vec{\zeta}_S \}$, and after tracing over the hidden qumode, we obtain the state of the visible mode,
\be\label{eq:rhov} \rho_v = \mathrm{Tr}_h \ket{\text{out}} \bra{\text{out}} \propto \mathrm{Tr}_h \mathcal{T} (\Vec{\zeta}) \ket{\psi_0} \bra{\psi_0} \mathcal{T}^\dagger (\Vec{\zeta}) \propto \mathcal{T}(\Vec{\zeta})  e^{-2\delta N}\mathcal{T}^\dagger (\Vec{\zeta})
\ee
where we used Eq.\ \eqref{eq:psi} and the fact that $\mathcal{T} (\Vec{\zeta})$ only acts on the visible qumode.
Evidently, this is a Hermitian operator. Hence we can define an effective Hamiltonian $H_{\text{eff}} (\Vec{\zeta})$ (a Hermitian operator) by
\be \rho_v = e^{-H_{\text{eff}} (\Vec{\zeta})} \ee
which depends on the set of parameters $\Vec{\zeta}$, showing that the output state of the visible qumode in the CVQBM can be viewed as a thermal state. The action of each non-unitary operator $\mathcal{T}_s (\Vec{\zeta}_s)$ can be thought of as a QITE step \cite{zoufal2021variational, yeter2022quantum}.
Thus, our quantum circuit \emph{defines} the Hamiltonian and its parameters to be trained.

Having designed the CVQBM circuit, we may proceed to use the machine to solve a variety of problems, such as generative learning tasks \cite{zoufal2021variational, kieferova2017tomography, wiebe2019generative}, discriminative tasks \cite{zoufal2021variational, amin2018quantum}, reinforcement learning \cite{crawford2016reinforcement}, quantum state tomography \cite{kieferova2017tomography}, and anomaly detection \cite{stein2023exploring, templin2023anomaly}. Here, we will concentrate on using the CVQBM as a continuous distribution generator. Our goal is to make the machine learn a continuous distribution and then use the trained machine to generate samples that belong to that distribution. This type of machine can be utilized as a synthetic data generation tool. There have been various attempts to use QBM as a generation tool, such as solving a toy problem of the phase of a state at 10 spatial points \cite{srivastava2023generative}, generating probability distributions for a state \cite{zoufal2021variational, kieferova2017tomography} and Bernoulli distribution \cite{amin2018quantum}. Generating a continuous distribution in a CV setting provides a natural advantage. 


We simulated the CVQBM using the Strawberry Fields software platform \cite{strawberryfields}. The quantum machine learning toolbox application is built on top of it, with
TensorFlow \cite{abadi2016tensorflow} features. The detailed architecture for our CVQBM is shown in Fig.~\ref{fig:QBMarch}. We started with two modes (visible and hidden), which were then entangled to prepare the two-mode initial state using the quantum gate $U_0$ depicted in Fig.~\ref{fig:U0det}. The visible mode was then acted upon with a series of non-unitary gates $\mathcal{T}_s (\vec{\zeta}_s)$ ($s=1,\dots, S$) whose detailed architecture is shown in Fig.\ \ref{fig:Udet}. The parameters of these gates are training parameters. At each step, an ancilla coherent mode was introduced, which was entangled with the visible mode by using Gaussian gates and then measured, thus projecting it onto the $n=1$ state. The success probability was calculated, ensuring that the setup corresponded to a reasonable rate for experimental implementation. Finally, we performed homodyne measurement on the visible mode.

In the training stage of our CVQBM, our objective was to minimize the distance between the target and output distributions. However, comparing two distributions at every training epoch is computationally intensive. We found it advantageous to work with density matrices. For a given set of data of probability distribution $P_{\mathrm{data}} (q)$, where $\int dq P_{\mathrm{data}} (q) =1$, we introduced the state
\be \ket{\Psi_{\mathrm{data}}} \approx \sum_{n=0}^{n_{\mathrm{max}}} a_n \ket{n} \ , \ \ a_n = \int dq \Psi_n (q) \sqrt{P_{\mathrm{data}} (q)} \ee
where $\Psi_n (q)$ is the $n$th state of a harmonic oscillator in the position representation given by Eq.\ \eqref{eq:psi}. This can be written in 
a density matrix form as
\be \rho_{\mathrm{data}} = \ket{\Psi_{\mathrm{data}}} \bra{\Psi_{\mathrm{data}}} \ , \ \ \rho_{\mathrm{data}, mn} = a_m a_n  \label{eq:ptod} \ee
which can be compared with the output state $\rho_v$ given in Eq.\ \eqref{eq:rhov}. The fidelity is defined as:
\begin{equation}
    F(\rho_{\mathrm{data}}, \rho_{v}) = \left( \textup{Tr} \sqrt{\sqrt{\rho_{\mathrm{data}}}\rho_{v} \sqrt{\rho_{\mathrm{data}}}} \right)^2
\end{equation}
Our goal was to maximize this fidelity, and hence minimize the cost:
\begin{equation}\label{eq:cost}
    C = 1 - F
\end{equation}
To this end, we used the Adam Optimizer. 

Once the CVQBM was trained, the next step was to generate more samples by loading the trained weights into the circuits. Since, ultimately, our goal is to generate the learned distribution, we had to extract the output from the circuit that corresponded to the output probability distribution,
\be P_v(q) = \braket{q\vert \rho_{v} \vert q} \ . \ee
Switching to eigenstates of the photon number, the output probability distribution can be written as
\be P(q) = \sum_{m,n} \rho_{v,mn} \Psi_m (q) \Psi_n (q) \ , \ \ \rho_{v,mn} = \bra{m} \rho_v \ket{n} \label{eqrhotop} \ee
where $\Psi_n (q)$ is given by Eq.~\eqref{eq:psi}. Although this is an infinite set of basis states, in practice we need to introduce a cutoff $n_{\text{max}}$ in the dimension of the Hilbert space.
Using this result, we can generate various samples from the trained CVQBM. 

A metric to compare the target with a generated distribution is provided by the Kuller-Leibler (KL) divergence,
\be\label{eq:KL} \textup{KL} (P_{\text{gen}}\| P_{\text{target}}) = \sum_{q} P_{\text{gen}}(q) \log  \frac{P_{\text{gen}}(q)}{P_{\text{target}}(q)}  \ee
which indicates how much the generated probability distribution, $P_{\text{gen}}$,  diverges from the target probability distribution, $P_{\text{target}}$,  and takes values between 0 and $\infty$, with the value of 0 when $P_{\text{gen}}(q)=P_{\text{target}}(q)$. We compared the two distributions and obtained how much the generated distribution diverged from the target distribution by calculating the KL divergence. 



\section{Classical Data Probability Distribution Generation} \label{sec:classicaldata}
Our CVQBM can be used for generating different kinds of distribution involving classical or quantum data. In this Section, we report on probability distributions from synthetic-aperture radar (SAR) images which are classical data. We used the SAR dataset available on the ICEYE website \cite{ICEYE}.

\begin{figure}[!tbp]
  \centering
  \subfloat[]{\includegraphics[scale=0.04]{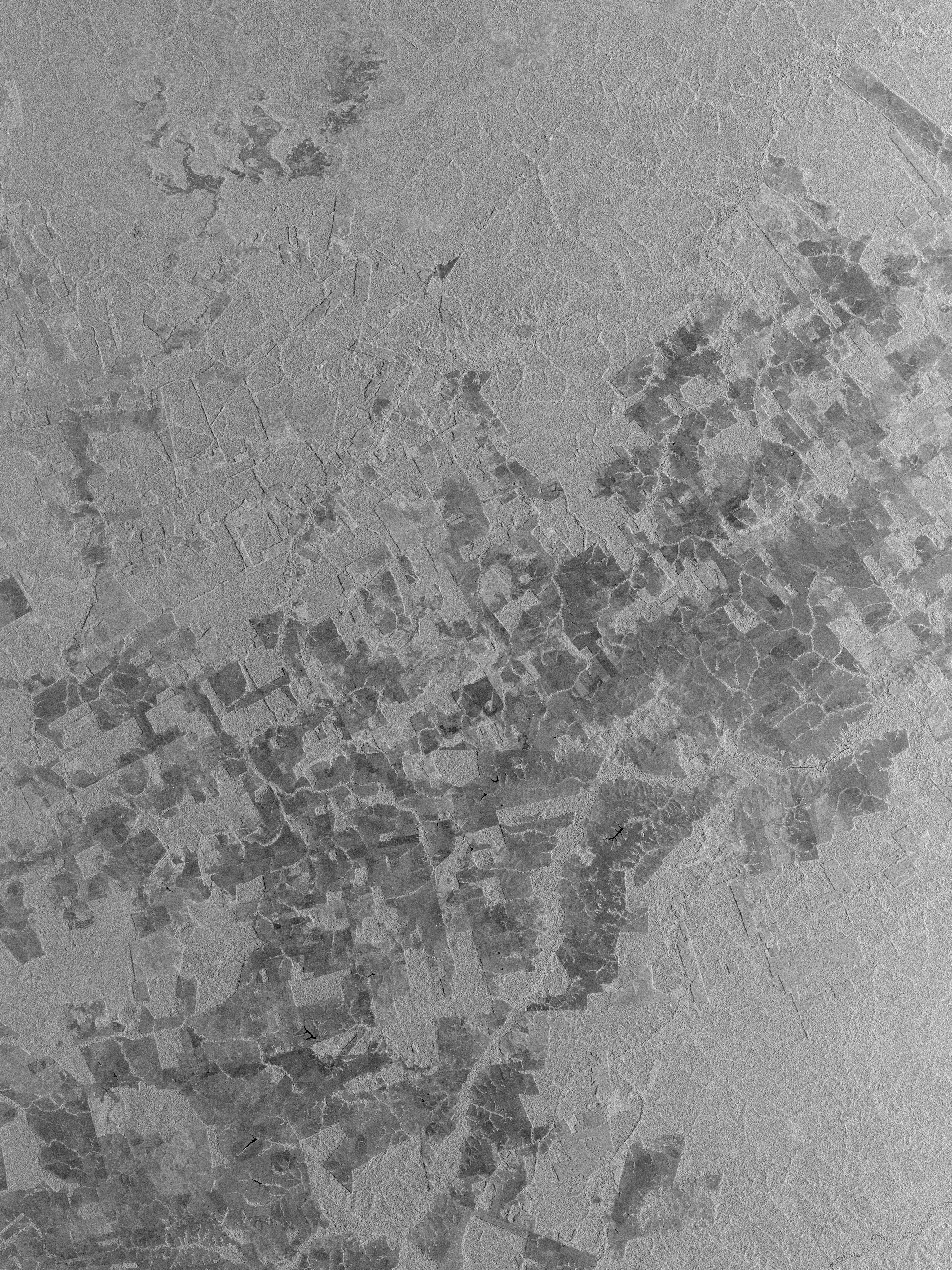}\label{fig:forest_f1}}
  \hfill
  \subfloat[]{\includegraphics[width=0.58 \textwidth]{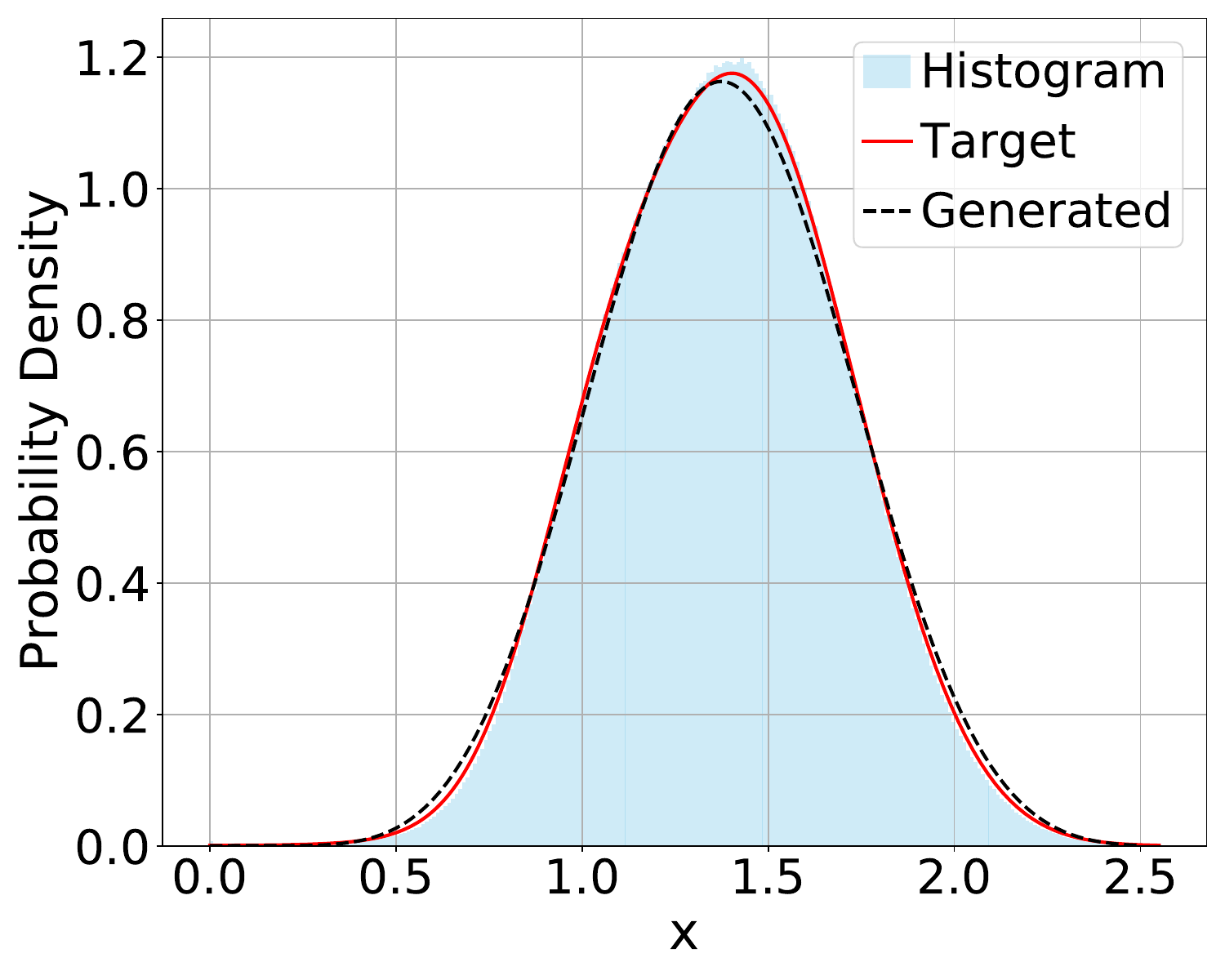}\label{fig:forest_f2}}
  \caption{Studying SAR image for forest monitoring in Brazil, \textbf{(a)} The original SAR image of the forest in Acre Brazil taken on 22 July 21. \textbf{(b)} The final plot corresponds to the forest monitoring. The blue histogram is created from the SAR image. The target distribution (red line) is fitted Gaussian on the histogram. Finally, the generated distribution (black dashed line) comes from the trained QBM, which has the KL divergence between the target and generated distribution is 0.006.}
\end{figure}

We selected a forest monitoring strip from Acre, Brazil. The dataset contains the time-series stack of strip SAR images acquired within 12 days in Single Look Complex (SLC) in HDF5 format and Ground Range Detected (GRD) in GeoTiff format. For simplicity, we worked with the PNG image in the dataset, shown in Fig.~\ref{fig:forest_f1}. The image belongs to the data collected on 22 July 2021. 

We analyzed the data by plotting the image intensity probability distribution as a function of the pixel intensity. To this end, we converted the image into a Python Numpy array, and then created the histogram for the intensity values. The histogram was divided into 256 bins (for each possible intensity value in an 8-bit grayscale image) and counted how many pixels fell into each bin. The range was set from 0 to 255, including all possible intensity values. The histogram was then normalized to yield a probability distribution. The bar graph was plotted to visualize the probability distribution of the pixel intensities. Next, we applied re-scaling to both plot axes to make the problem suitable for a CV quantum computer. By doing so, the range of axes became approachable, and normalization was maintained. Because of the limitations of the simulation, it was hard to use the discrete data as this would require a higher cutoff value. Hence, the next step was to make the distribution continuous, and for that, we used the \textit{gaussian kde} from the SciPy library \cite{SciPy} of Python. Once we obtained the distribution, we used Eq.~\eqref{eq:ptod} to deduce the density matrix. Using the density matrix for cost calculation provided a computational advantage over using the distribution directly. 

As discussed in Section \ref{section:methods}, the cost function depends on the fidelity of the circuit output density matrix compared with the target density matrix. For an optimal entangled initial state (Eq.\ \eqref{eq:32}), we set $\delta = 1.5$. We used the Adam optimizer with the learning schedule of exponential decay to minimize the cost. The initial learning rate was 0.05; the decay steps were 100, with a decay rate of 0.96. The number of steps in this process was $S=2$. After training for 100 epochs with a batch size of 32 and a cutoff of $n_{\text{max}} =12$, the best fidelity we obtained was 99.8\%, corresponding to the cost value of 0.002. The low cost made us confident that the model had been trained well. 
\begin{figure}[!tbp]
  \centering
  \subfloat[]{\includegraphics[width=0.42\textwidth]{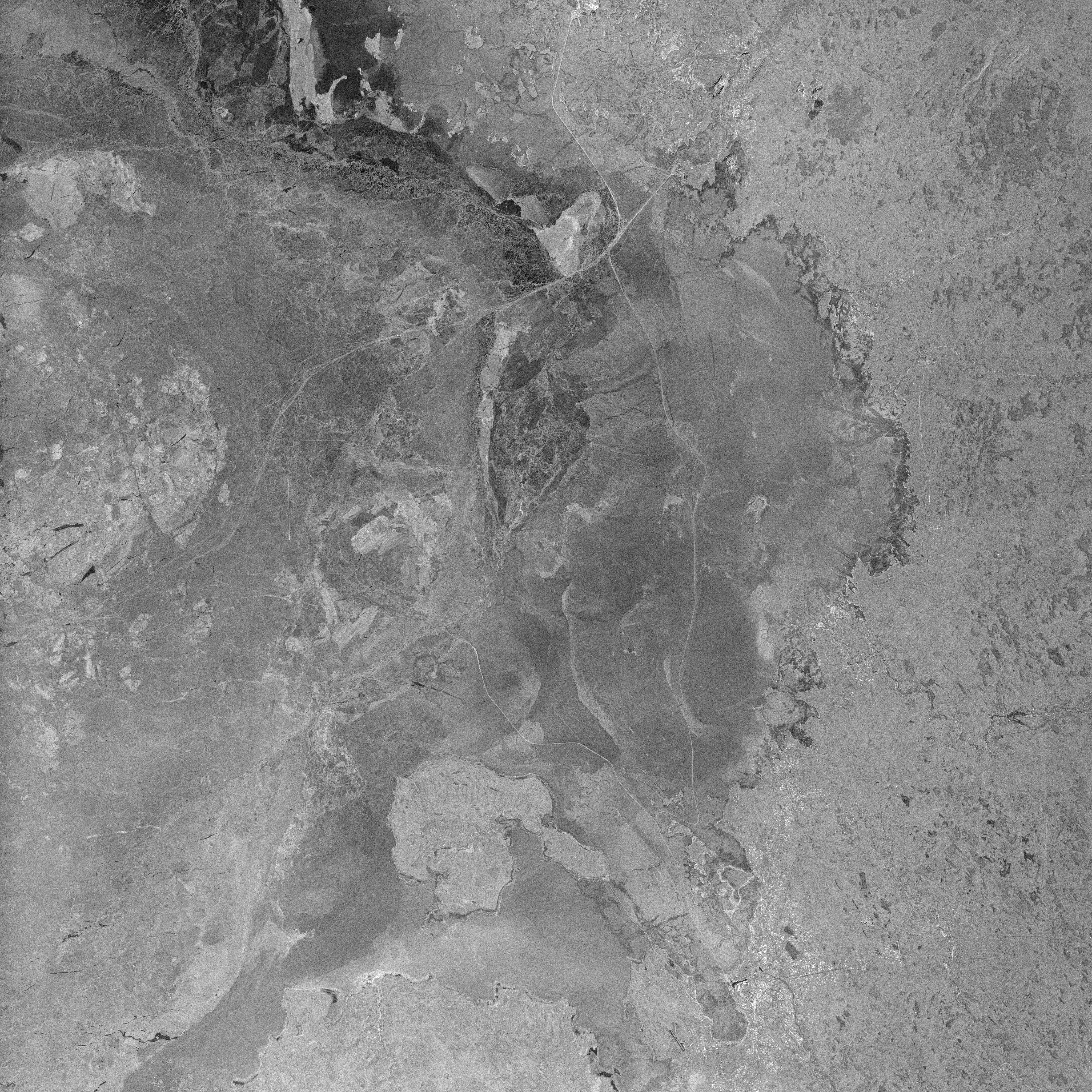} \label{fig:originalGulf}} 
  \hfill
  \subfloat[]{\includegraphics[width=0.5 \textwidth]{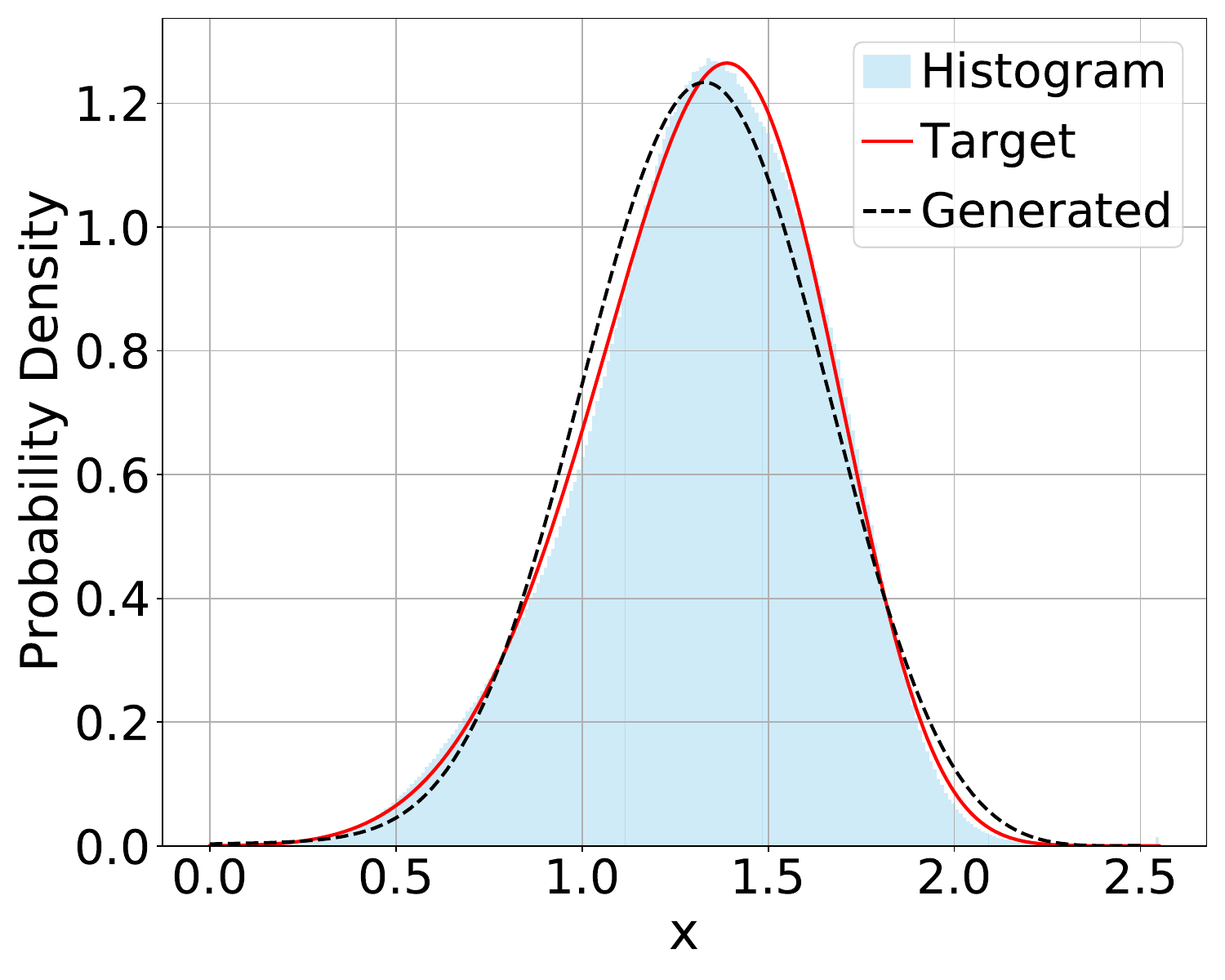} \label{fig:generatedocean}}
  \caption{Studying the SAR image for sea ice monitoring, \textbf{(a)} the original SAR image of Gulf of Bothnia, taken on 25 Feb 2022, \textbf{(b)} The final plot corresponds to the sea monitoring. The blue histogram is created from the SAR image. The
target distribution (red line) is fitted Gaussian on the histogram. Finally, the generated
distribution (black dashed line) comes from the trained QBM, which has the KL divergence between the target and generated distribution is 0.013} 
\end{figure}

After training, the model can be used to generate more data corresponding to the target distribution. For this process, we loaded the trained weights and ran the circuit several times to get samples for the density matrix. We generated 256 samples corresponding to the original shape of the data. Then using Eq.~\eqref{eqrhotop}, we generated the distribution. The generated and target distributions, as well as the original discrete probability distribution, are shown in Fig.~\ref{fig:forest_f2}. Notice that the generated distribution lies on top of the target distribution, showing that we achieved high fidelity while training.  We also calculated the KL divergence between the generated and target distributions, which came out to be 0.006, showing that the generated distribution is close to the target distribution. 


Generating the distribution for the forest was not hard, as we could fit the probability distribution with a Gaussian. A more challenging example we studied was provided by an image that belonged to the Gulf of Bothnia between the west coast of Finland and the east coast of Sweden \cite{ICEYE}, shown in Fig.~\ref{fig:originalGulf}. The dataset had a pair of scan mode images, which covered an area of 10,000 km$^2$ in the Gulf of Bothnia. Again, we randomly picked one of the images and plotted the probability distribution vs.\ pixel intensity. This probability distribution was  best fitted with a 
Weibull probability distribution:
\be
W(x; \lambda, k) = \frac{k}{\lambda} \left( \frac{x}{\lambda} \right)^{k-1}
\ee
where $x \geq 0$, $k>0$ is the shape parameter, and $\lambda>0$ is the scale parameter of the distribution. For this case, the fitted Weibull distribution had $\lambda = 161.2$ and $k=5.4$. 


We adopted the same setup as in the forest case. First, we found the density matrix corresponding to the probability distribution. Then, the cost function was defined in terms of the fidelity between the target density matrix and the circuit output density matrix. Again, we used the Adam optimizer with the same learning schedule as the previous case. The number of steps used in this process was $S=2$, with a cutoff dimension of $n_{\text{max}} = 15$. After training for 150 epochs with a batch size of 32, the best fidelity we obtained was 98.9\%. The final plot is shown in Fig.~\ref{fig:generatedocean}. The generated distribution in the dashed black line matches the target Weibull distribution (shown in red solid line). 
We obtained KL divergence of 0.013.

\begin{figure}[!tbp]
  \centering
  \subfloat[]{\includegraphics[width=0.32\textwidth]{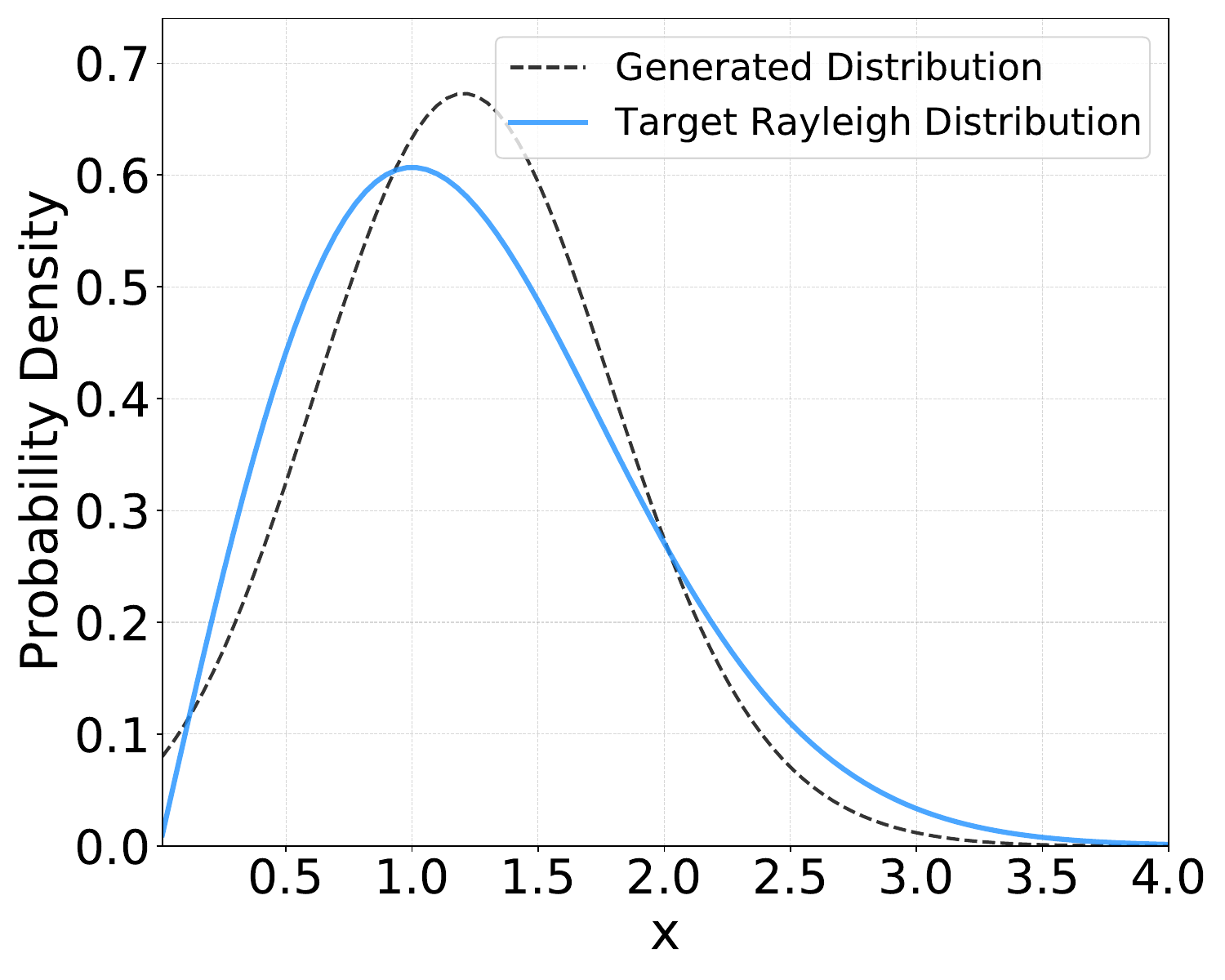}\label{fig:Ray}}
  \hfill
  \subfloat[]{\includegraphics[width=0.32 \textwidth]{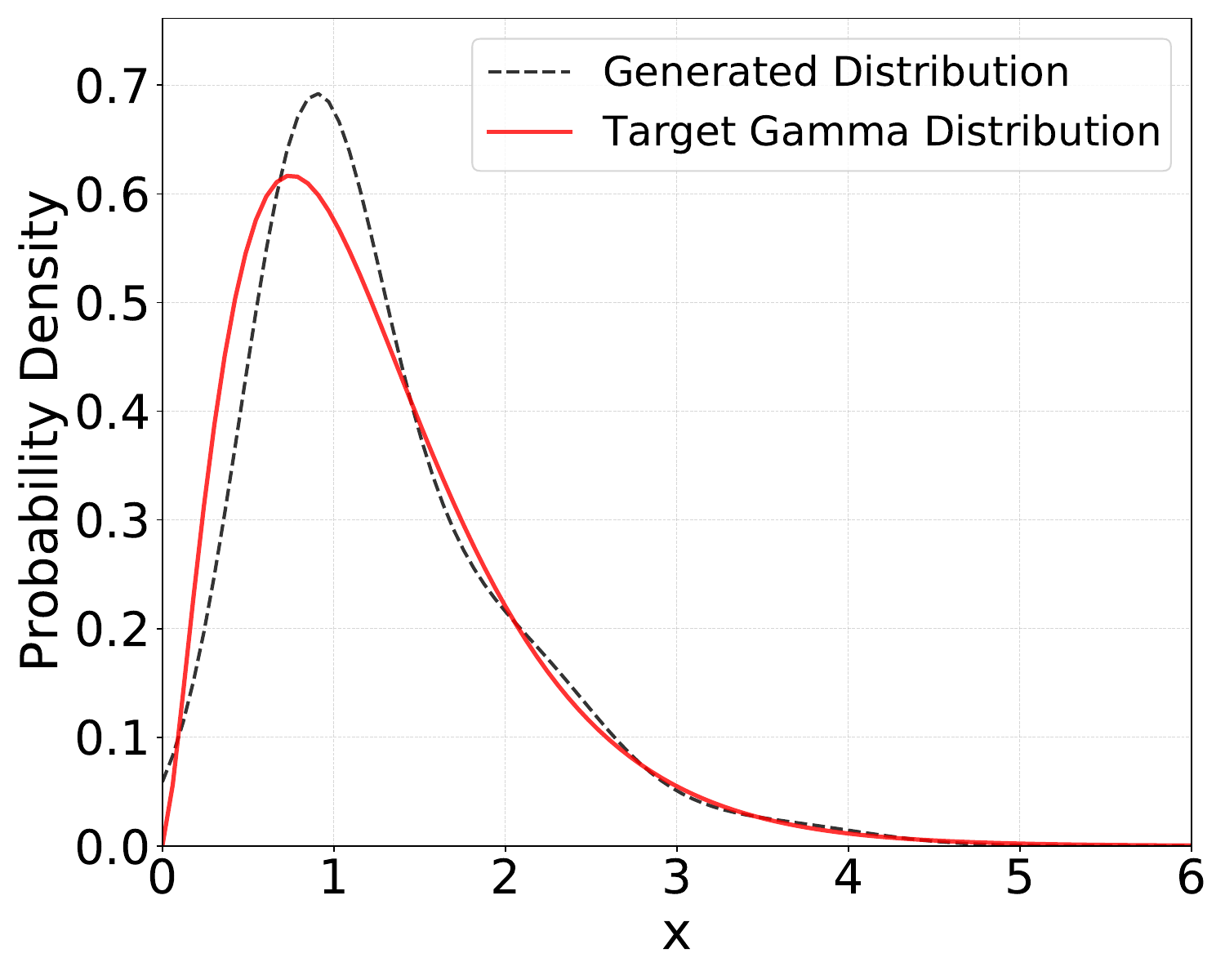}\label{fig:gamma}}
    \hfill
  \subfloat[]{\includegraphics[width=0.32 \textwidth]{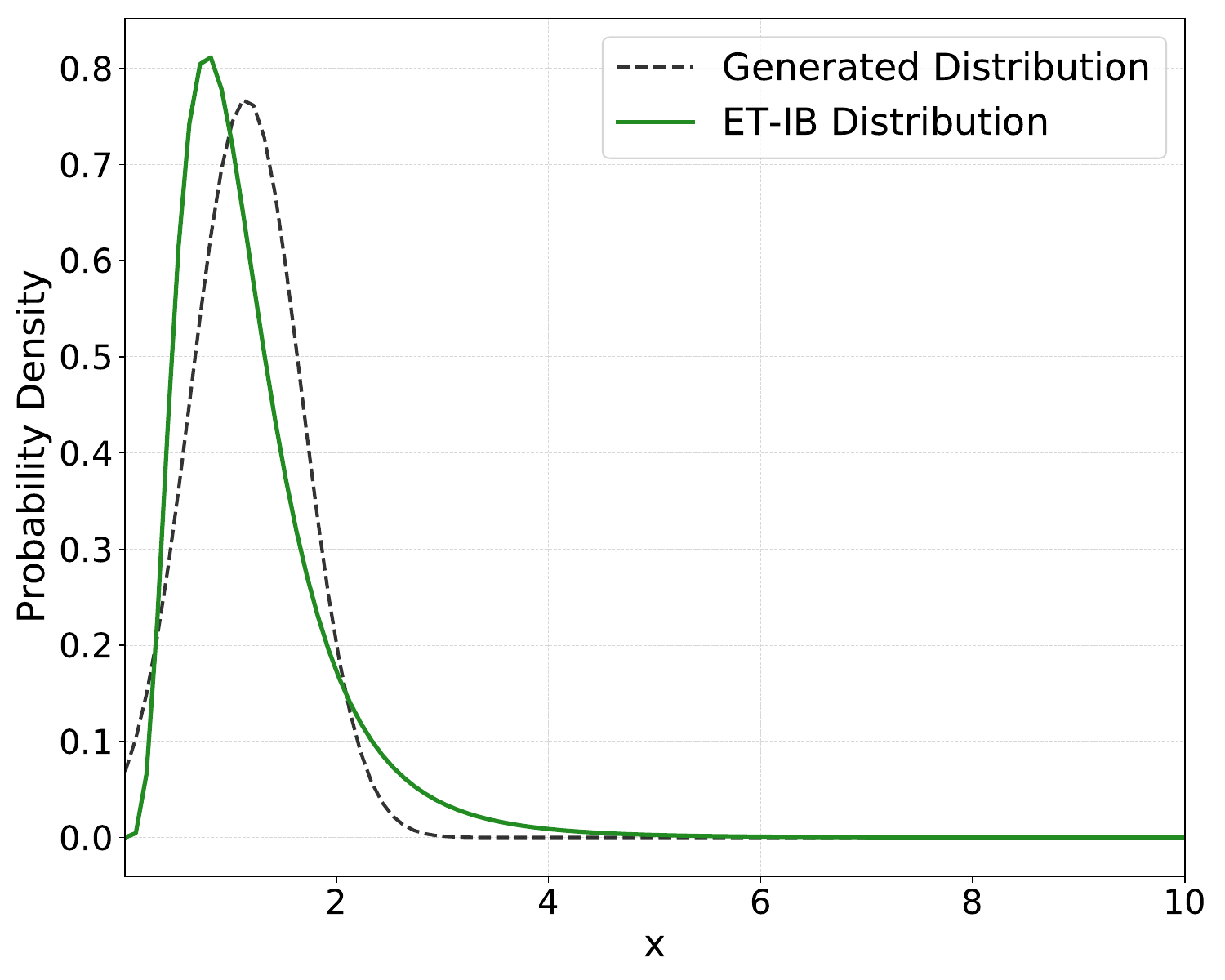}\label{fig:etib}}
  \caption{Generating continuous probability distributions implementing some commonly used distribution \textbf{(a)} Rayleigh, \textbf{(b)} Gamma and \textbf{(c)} ET-IB distribution} \label{fig:differentdist}
\end{figure}

Our model works with various probability distributions requiring an increasing amount of computational resources as their complexity increases. We demonstrated this with several distributions commonly associated with real-world data, including Rayleigh and Gamma distributions which are often used for SAR and medical images, like ultrasound and PET scans, as well as exponentiated transmuted-inverted beta (ET-IB) distributions that were recently shown \cite{sagrillo2022new} to outperform (in terms of goodness-of-fit measures) its competitors when it comes to fitting SAR images. We studied the performance of our model on these distributions as follows:
\begin{itemize}
  \item \underline{Rayleigh distribution}, defined as
  \be R(x;\sigma) = \frac{x}{\sigma^2} \textup{exp}\left(-\frac{x^2}{2\sigma^2}\right), \ \ x \geq 0 \ee where $\sigma$ is the scaling parameter.  In Fig.~\ref{fig:Ray}, we chose $\sigma =1$ and obtained a fidelity of 96.1\% after training and KL divergence of 0.03. We set $\delta = 2.5$ for an optimal entangled initial state (Eq.\ \eqref{eq:32}). A real-world application where the Rayleigh distribution is observed is the wind speed in a city, studied over a time period in Ref.\ \cite{serban2020assessment}. It was found that $\sigma=4$ (see Table \ref{tab:dist}). 
  \item \underline{Gamma distribution}, defined as
  \be G(x;k, \theta) =  \frac{x^{k - 1}}{\Gamma(k) \theta^k} \textup{exp}\left( -\frac{x}{\theta} \right), \ \ x\geq 0\ee where $k$ is the shape parameter, and $\theta$ is the scaling parameter. In Fig.~\ref{fig:gamma}, we chose $k =2.5$ and $\theta =0.5$ and obtained a fidelity of 98.1\% after training and  KL divergence of 0.083. We set $\delta = 2.5$ for an optimal entangled initial state (Eq.\ \eqref{eq:32}). There are many applications of the Gamma distribution. As an example, to study the ultrasound images displaying breast cancer, various parameter sets of distribution are used: $(k, \theta) \in \{ (1,2), (2,0.5), (2,1), (2,1), (2,4), (4,2) \}$ \cite{borlinhas2019gamma} (see Table \ref{tab:dist}).    
  
  \item \underline{ET-IB distribution}, defined as 
  \bea f(x;\alpha, \beta, \lambda, \phi) &=& \frac{\phi x^{\alpha-1} I_{x/(1+x)} (\alpha, \beta)^{\phi-1}}{(1+x)^{\alpha + \beta} B(\alpha, \beta)} \nonumber\\
  && \times \left( 1+\lambda -2\lambda I_{x/(1+x)} (\alpha, \beta) \right) \left( 1 + \lambda - \lambda I_{x/(1+x)} (\alpha, \beta) \right)^{\phi-1} \ ,\ \  \ \ \ 
  \eea
  where 
  \begin{equation}
      I_z (\alpha, \beta) = \frac{1}{B(\alpha, \beta)} \int_0^z t^{\alpha-1} (1-t)^{\beta -1} dt \ , \ \
      B(\alpha,\beta) = \frac{\Gamma(\alpha)\Gamma(\beta)}{\Gamma(\alpha+\beta)}
  \end{equation}
  The parameters are constrained as $\alpha, \beta, \phi>0$ and $|\lambda|<1$. For Fig.~\ref{fig:etib}, we chose the four parameters $(\alpha, \beta, \lambda, \phi) = (4.0, 5.0, 0.1, 2.0)$, and obtained a fidelity of 91.4\% after training, and KL divergence of 0.229. We set $\delta = 4.0$ for an optimal entangled initial state (Eq.\ \eqref{eq:32}). In Ref.\ \cite{sagrillo2022new}, the Monte Carlo technique was used to perform parameter estimation for SAR images using different terrains. For example, the parameters $(\alpha, \beta, \lambda, \phi)$ were found to be (6.3, 36.2, -0.9, 0.4) for forest images, (3, 37.5, 0.8, 3.7) for the ocean, and (0.4, 5.3, 0.9, 66.9) for urban (see Table \ref{tab:dist}).  
\end{itemize}
We used the same Adam optimizer with the learning schedule for all cases as before. The number of steps was $S=2$, the cutoff dimension was $n_{\text{max}} =15$, and the number of epochs was 100. The final plots for all three distributions are shown in Fig.~\ref{fig:differentdist}. The colored solid lines correspond to the different target distributions, and the black dashed lines correspond to the generated distribution. In all three cases, we achieved high fidelities. An important metric indicating how well the generated distribution compares to the target distribution is the KL divergence. This is evident in the plots, with the Rayleigh distribution having a lower KL divergence and better fitting the histogram compared to the ET-IB distribution, which has a higher value of KL divergence. 
As discussed above, these distributions have numerous real-world applications. In Table \ref{tab:dist}, we compare the distribution parameters used in this work with some of the real-world applications. The parameters used for the Gamma distribution are similar to those used in studying ultrasound images of breast cancer. However, this was not the case for Rayleigh and ET-IB distributions because they are steeper and require a higher cutoff dimension in simulations. The study of these real-world applications would require high-performance computers, or more efficient simulations, or an actual CV quantum computer which is not currently available.

\begin{table}[]
\centering
    \caption{Description of parameters used for different distributions and their comparison with real-world applications.}
\begin{tabular}{@{}cccccc@{}}
\toprule
\textbf{Names} & \textbf{Parameters} & \begin{tabular}[c]{@{}l@{}}\textbf{Parameters used}\\ \textbf{in this work}\end{tabular} & \textbf{Applications }& \begin{tabular}[c]{@{}l@{}}\textbf{Parameters used}\\ \textbf{in the applications}\end{tabular}                                           \\ 
\midrule
Rayleigh &  $\sigma$   & 1           & Wind speeds \cite{serban2020assessment} & 4   \\
\midrule
Gamma    &   (k, $\phi$)  & (2.5, 0.5)  & \begin{tabular}[c]{@{}l@{}}Breast cancer\\ ultrasound images \cite{borlinhas2019gamma}\end{tabular}  & \begin{tabular}[c]{@{}l@{}}(1, 2), (2, 0.5), (2, 1),\\ (2, 4), (4, 2)\end{tabular}      \\
\midrule
ET-IB    &  ($\alpha$, $\beta$, $\lambda$, $\theta$) & (4, 5, 0.1, 2) & \begin{tabular}[c]{@{}l@{}}SAR images \cite{sagrillo2022new}:\\ forest\\ ocean\\ urban\end{tabular} & \begin{tabular}[c]{@{}l@{}}\\ (6.3, 36.2, -0.9, 0.4)\\ (3.0, 37.5, 0.8, 3.7)\\ (0.4, 5.3, 0.9, 66.9)\end{tabular} \\ \bottomrule
\end{tabular} \label{tab:dist}
\end{table}

\section{Quantum Data Probability Distribution Generation} \label{sec:quantumdata}


\begin{figure}[ht!]
    \centering
    \includegraphics[trim={1cm 12cm 13cm 4cm}, clip,width=0.7\textwidth]{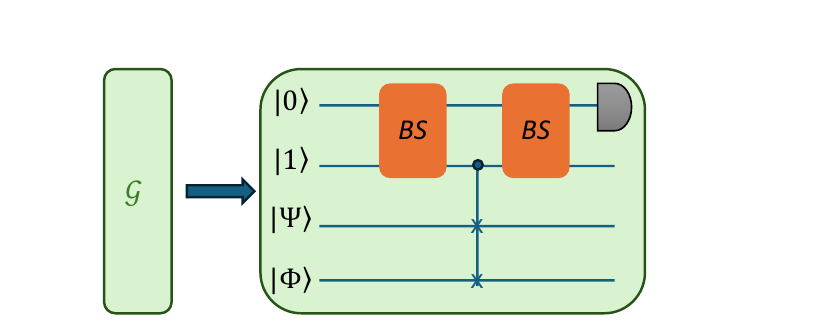}
    \caption{Gadget ($\mathcal{G}$) for quantum data to be attached to the CVQBM during training. $\ket{\Psi}$ is the target state and $\ket{\Phi}$ is the output of the CVQBM. The first two lines represent ancilla qumodes initialized in states $\ket{0}$ and $\ket{1}$, respectively.}
    \label{fig:qgadget}
\end{figure}

Next, we apply our CVQBM to quantum data. We consider two cases: a Gaussian distribution and a non-Gaussian distribution (cat state) both of which are generated approximately by a quantum neural network
(QNN) \cite{bangar2023experimentally}. 

To handle quantum data, we need an additional gadget, $\mathcal{G}$, to attach to the CVQBM shown in Figure \ref{fig:qgadget}. The state
 $\ket{\Psi}$ represents the target quantum state, generated by a quantum circuit, and $\ket{\Phi}$ represents the output state of the visible mode of our CVQBM. We introduce two control qumodes in the states $\ket{0}$ and $\ket{1}$, respectively, and put them in the state 
\be \ket{\text{ctrl}} = \frac{1}{\sqrt{2}} ( \ket{01} +  \ket{10} ) \ee 
with the aid of a 50:50 beam splitter. Then we apply the three-mode CSWAP gate 
\be U_{\text{CSWAP}} \ket{\text{ctrl}} \ket{\Psi} \ket{\Phi} = \frac{1}{\sqrt{2}} ( \ket{01} \ket{\Psi} \ket{\Phi} +  \ket{10} \ket{\Phi} \ket{\Psi} ) \ee 
which can be implemented with the aid of a non-Gaussian Kerr gate \cite{PhysRevLett.118.080501,marshall2015repeat}. After channeling the beams through another 50:50 beam splitter, we obtain
\be U_{\text{BS}} \cdot U_{\text{CSWAP}} \ket{\text{ctrl}} \ket{\Psi} \ket{\Phi} = \frac{1}{2} ( \ket{01}  (\ket{\Psi} \ket{\Phi} - \ket{\Phi} \ket{\Psi}) +  \ket{10}  (\ket{\Psi} \ket{\Phi} + \ket{\Phi} \ket{\Psi}) )~. \ee 
Measurement of the first control qumode yields probabilities $P_i$ ($i=0,1$) with
\be P_1 - P_0 = |\braket{\Phi | \Psi}|^2~, \ee
which is the fidelity to be used in the definition of the cost function \eqref{eq:cost}.

\begin{figure}[!tbp]
  \centering
  \subfloat[]{\includegraphics[width=0.48\textwidth]{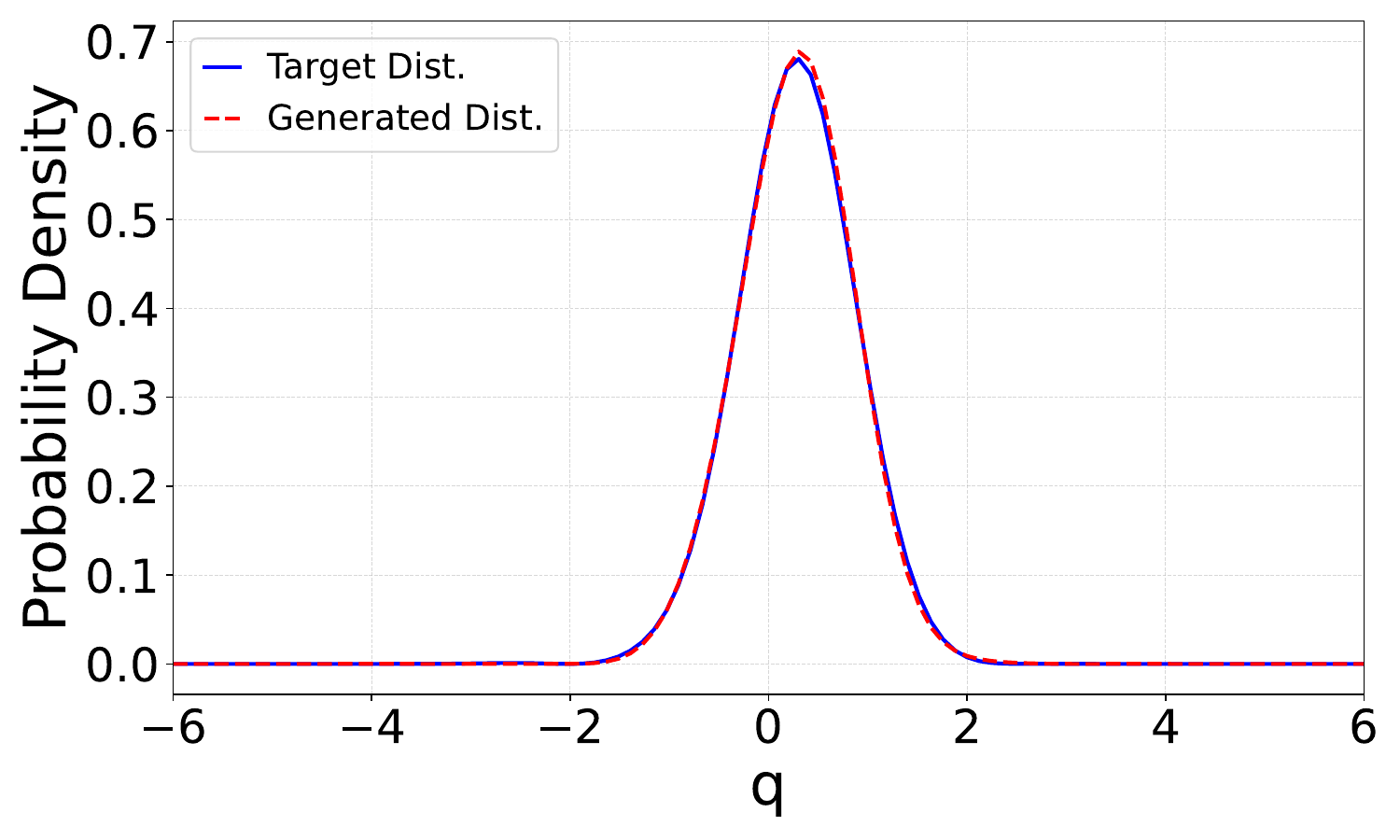}}
  \hfill
  \subfloat[]{\includegraphics[width=0.48 \textwidth]{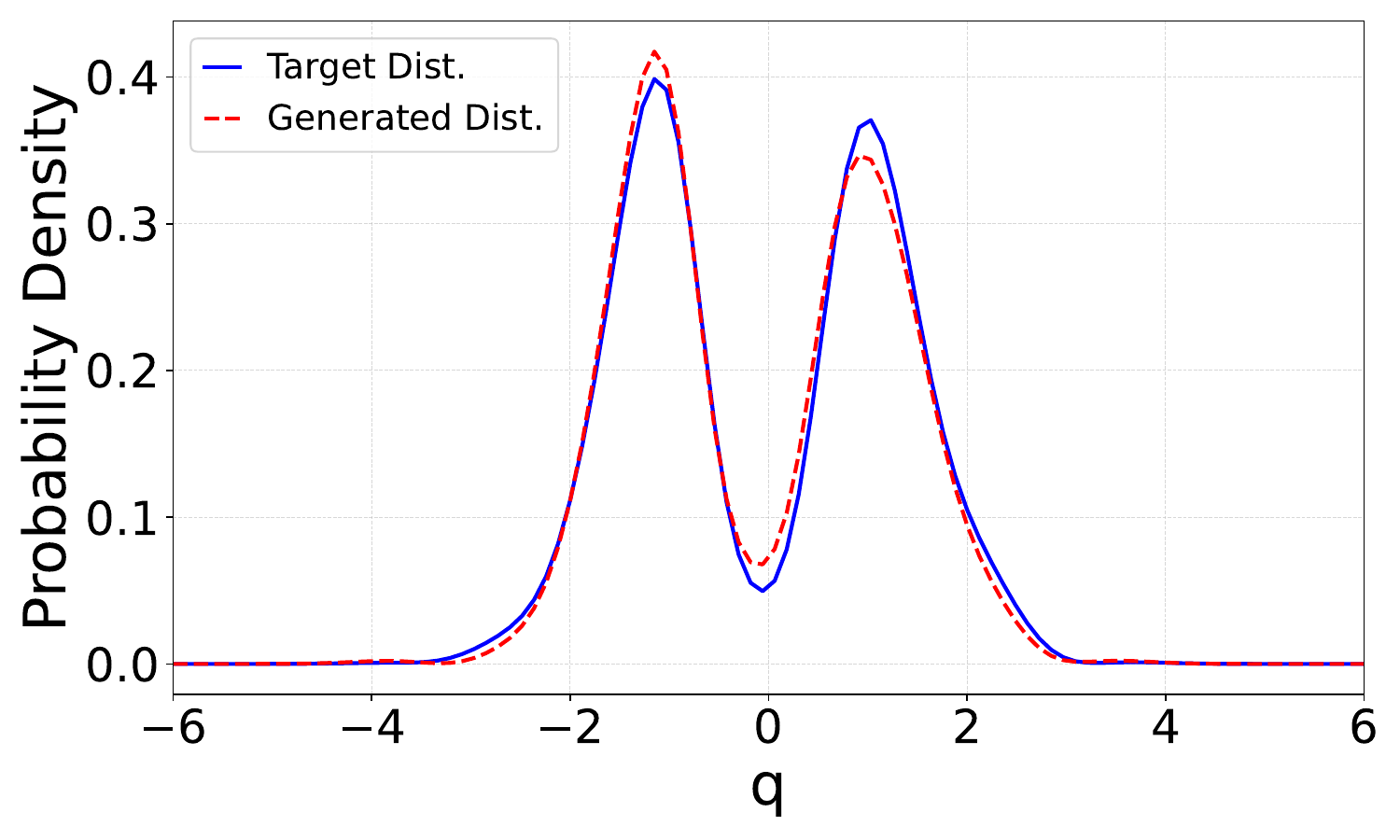}}
  \caption{Generating continuous probability distributions from quantum data: (a) Gaussian distribution and (b) non-Gaussian distribution resulted from a cat state preparation. The target (solid blue lines) and generated (dashed red lines) distributions match well.} \label{fig:quantumdata}
\end{figure}

For the preparation of quantum data, we relied on our previous work on experimentally realizable CV QNNs \cite{bangar2023experimentally}. More specifically, we used the circuits for state preparation to prepare a Gaussian and non-Gaussian state approximately. 

For a Gaussian state, we constructed a target Gaussian state for our CVQBM from a single qumode in the vaccum state on which we applied a squeezing gate with a squeezing of 1.73 dB and a displacement gate with a displacement of parameter $\alpha =0.2$. The resulting Gaussian state had a mean of 0.083 and a standard deviation of 0.145. We trained the CVQBM circuit with 1 step of our QITE algorithm and an initial entanglement of $\delta = 1.5$. Our trained CVQBM circuit attained a fidelity of 99.55\% resulting in a Gaussian state with mean 0.083 and standard deviation 0.147.

For the non-Gaussian state, our target quantum state was created using an 8-layer CV QNN \cite{bangar2023experimentally} which aimed to approximate an even squeezed cat state with a displacement magnitude of 1.2 and squeezing of 2.60 dB. It achieved this with a fidelity of 96.89\%. Our CVQBM, in turn, learned this non-Gaussian state with a fidelity of 98.22\%, utilizing an architecture that contained Gaussian gates only and an ancillary mode on which we performed a projective measurement, utilizing 3 steps of QITE and entangling parameter $\delta = 2.3$.

In the CV QNN simulation, we used a cutoff dimension of 10 and optimized the fidelity of the target state using the Nelder-Mead optimizer. For the CVQBMs, we used the Adam optimizer with a learning scheduler. The initial learning rate was 0.04 for both Gaussian and non-Gaussian data and both the Gaussian and the non-Gaussian states were trained for 400 epochs with a cutoff of 10.

The target and generated probability distributions are shown in Fig.\ \ref{fig:quantumdata}. The solid blue lines correspond to the target distributions, whereas the dashed red lines correspond to the generated distributions. The respective distributions match well, with KL divergence of 0.002 and 0.025 for the Gaussian and non-Gaussian (cat state) cases, respectively.

\section{Experimental Feasibility} \label{sec:exp}
A main goal in developing novel quantum algorithms is achieving experimental feasibility with currently available technology. In this Section, we discuss the experimental feasibility of the gates in our proposed CVQBM, the success probability, which is an important metric, since our CVQBM relies on non-deterministic processes (projective measurements on ancilla qumodes), and the effects of noise. 

For experimental feasibility, we restricted the parameters of the gates during training within an experimentally attainable range with current technology. 
This is especially challenging when dealing with real-world problems like studying SAR images on quantum optical hardware. The range of parameters for the gate set that were used in our case studies are shown on Table \ref{tab:1}. In addition to these gates, the CX gate ($e^{i\kappa q\otimes p}$), can be decomposed into two single-mode squeezers and beam splitters. Particularly, achieving high squeezing rates is one of the most challenging tasks in photonic quantum computing. Experiments are being conducted to achieve higher squeezing values. One of the highest squeezing value (15 dB) was reported in Ref.\ \cite{vahlbruch2016detection}. Hence, the squeezing parameters required to realize the use cases studied in this paper are within experimental range. 

\begin{table}[h]
    \centering
    \caption{Range of trained parameters of gates in the CVQBM circuit}
    \setlength{\tabcolsep}{10pt} 
    \renewcommand{\arraystretch}{1.4} 
    \label{tab:1}
    \begin{tabular}{@{}cccccccc@{}}
        \toprule
        & \textbf{Gates} & \textbf{Parameter} & \multicolumn{1}{c}{\begin{tabular}[c]{@{}c@{}}\textbf{Classical case}\\ \textbf{Forest}\end{tabular}} & \multicolumn{1}{c}{\begin{tabular}[c]{@{}c@{}}\textbf{Quantum case}\\ \textbf{Cat state}\end{tabular}} \\
        \midrule
        & Squeeze Gate & r (in dB) & 2.60 - 5.21 & 6.05 - 15.41\\
        & Displacement Gate & $\alpha$ & 0.8 - 1.3 & 0.01 - 0.51\\
        & $\text{CX}$ & $\kappa$ & 0.08 - 1.7 & 0.01 - 0.45\\
        \bottomrule
    \end{tabular}
\end{table}

Our CVQBM design is best suited for photonic quantum hardware. Unfortunately, currently available photonic quantum devices have limitations that prevent us from realizing our CVQBM with cloud-accessed photonic quantum hardware on chip. For example, the Xanadu X-series \cite{arrazola2021quantum}, which is a cloud-accessed quantum chip that includes squeezers, beam splitters, and interferometers cannot accommodate our CVQBM mainly due to the lack of online squeezers (the parameters of the available offline squeezers are fixed). The limitations of current photonic quantum hardware highlight the importance of co-designing quantum hardware, software, and algorithms. We leave the experimental realization of our CVQBM model on photonic quantum hardware to future work.

Turning to the success probability of our probabilistic algorithms, we note that the odds of a full pass-through of our CVQBM circuit is dictated by the probability of successfully measuring the required number of photons in the ancilla qumodes. Since we are post-selecting runs with a specific number of photons measured, the overall success probability is a product of success probabilities at each step of the QITE algorithm. 
If $p_s$ represents the probability of measuring 0 photons at the $s-$th QITE step, then the overall success probability is given by
$P_{\textrm{success}} = p_1  p_2 \cdots p_S$,
where $S$ is the total number of steps in the QITE algorithm. This multiplicative nature arises because the success of each step is conditional on the success of the previous steps. Therefore, to achieve a successful pass-through of the entire circuit, each individual step must succeed, leading to the cumulative success probability described above.

We studied the success probability of generating the forest and sea SAR images using our CVQBM model. For the parameters indicated in Section \ref{sec:classicaldata}, we obtained a success probability of 16.3\% (27.2\%) with fidelity 99.7\% (98.7\%) for the forest (sea) SAR images. It is possible to increase success probability at the expense of fidelity. For forest (sea) data, we obtained a success probability of 99.2\% (99.7) with a fidelity of 95.1\% (94.6\%).
Similar results were obtained for the other distributions we studied. For the Rayleigh, Gamma, and ET-IB distributions, we achieved fidelity of 96.2\%, 98.1\%, and 91.4\% at a success rate of 50.5\%, 1.6\%, and 49\%, respectively. Similarly, in the case of quantum data, we were able to train models with fidelity of 95.0\%  (98.2\%) at a success rate of 99.9\% (21.3\%) for Gaussian (non-Gaussian) states. Results on achieved fidelity and corresponding success rates are shown in Table \ref{tab:2}.

\begin{figure}[ht!]
  \centering
  \subfloat[]{\includegraphics[width=0.4\textwidth]{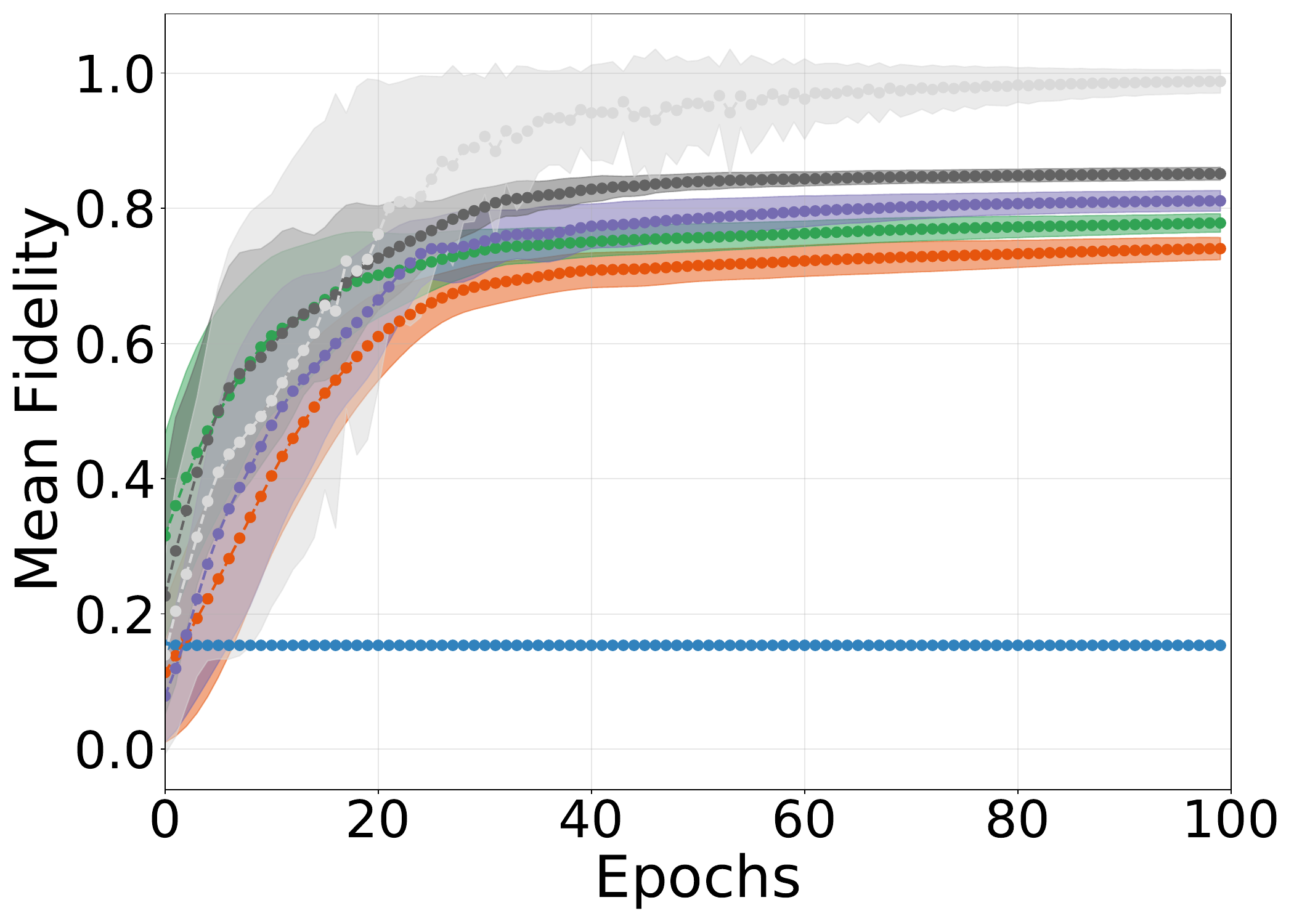}}
  \hfill
  \subfloat[]{\includegraphics[width=0.55\textwidth]{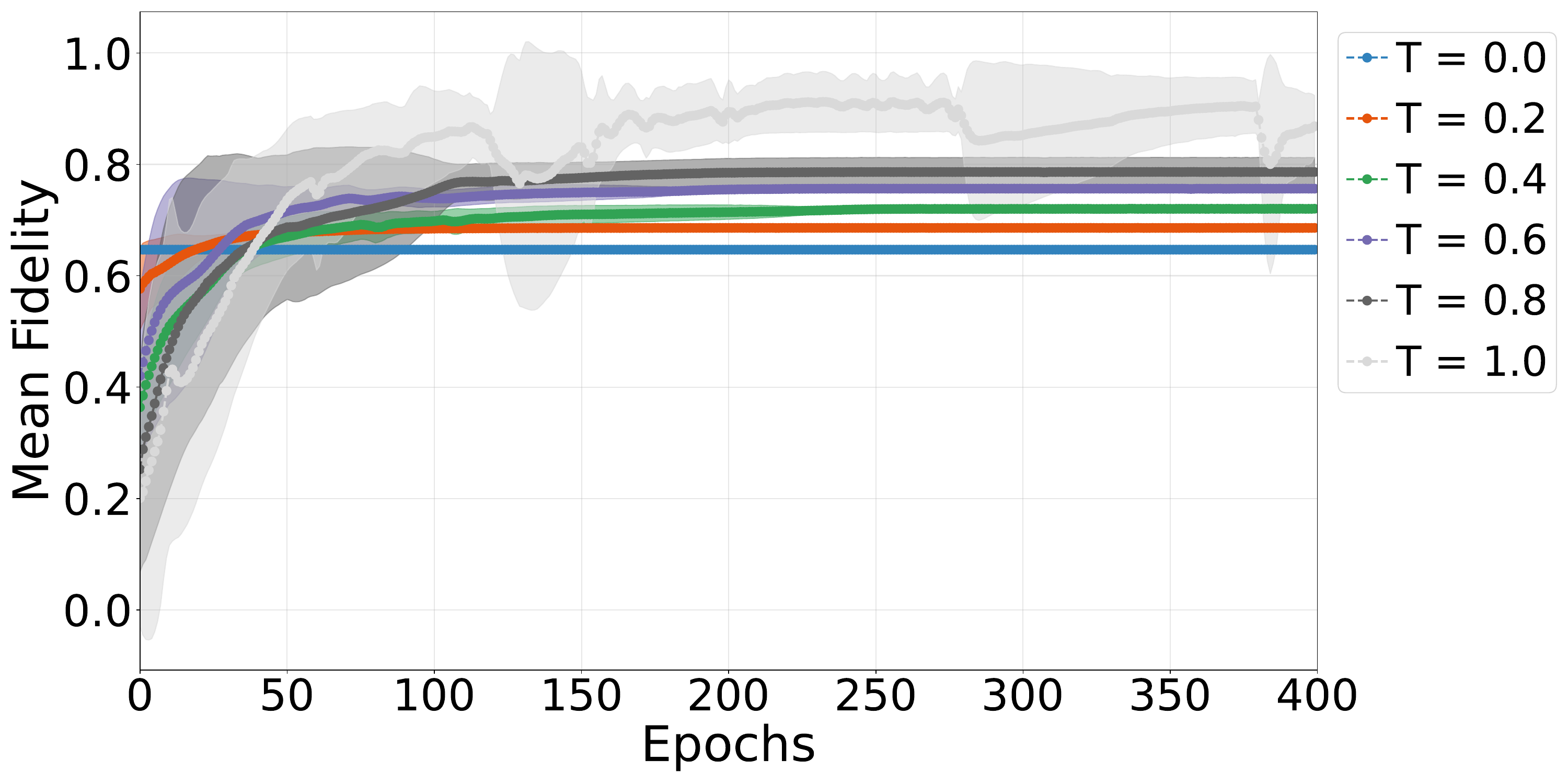}}
  \caption{Studying various distributions under the loss channel. Each loss parameter ($T$) run was done 5 times to get the mean cost values and the filled region around it represents the standard deviation of the fidelity. \textbf{(a)} Forest SAR images, and \textbf{(b)} Cat state distributions under the noise loss channel} \label{fig:addingnoise}
\end{figure}

\begin{table}[h!]
    \centering
    \caption{Achieved fidelity and success rates in use cases}
    \setlength{\tabcolsep}{10pt} 
    \renewcommand{\arraystretch}{1.4} 
    \label{tab:2}
    \begin{tabular}{@{}cccccccc@{}}
        \toprule
        & \textbf{Use case} & \textbf{Fidelity} & \textbf{Success Rate} & \textbf{} \\
        \midrule
        & Forest SAR Images & 99.7\%  & 16.3\% & \\
        &  & 95.1\%  & 99.2\% & \\
        & Sea SAR Images & 98.7\% & 27.2\% & \\
        &  & 94.6\%  & 98.7\% & \\
        & Rayleigh Distribution & 96.2\% & 50.5\% & \\
        & Gamma Distribution & 98.1\% & 1.6\% & \\
        & ET-IB Distribution & 91.4\% & 49\% & \\
        & Quantum Gaussian State & 95.0\% & 99.9\% & \\
        & Quantum Non-Gaussian (Cat) State & 98.2\% & 21.3\% & \\
        \bottomrule
    \end{tabular}
\end{table}

Finally, we added noise to our model to assess the robustness of the proposed CVQBM. We chose the simple noise model, loss channel, available in Strawberry Fields. This channel couples a given qumode in the circuit, with annihilation operator $\hat{a}$, to an unwanted qumode representing noise, with annihilation operator $\hat{b}$, prepared in the vacuum state. Thus,
\begin{equation}
\hat{a} \to \sqrt{T}\hat{a}+\sqrt{1-T}\hat{b}~,
\end{equation}
where $T \in [0,1]$ is the loss parameter, also known as energy transmissivity. $T=1$ represents the noiseless case (identity map) and for $T=0$, the state is mapped to the vacuum state. 
We introduced loss channels for the various qumodes in our circuits. 
\begin{figure}[ht!]
    \centering
    \includegraphics[trim={0cm 0cm 0cm 0cm}, clip,width=0.7\textwidth]{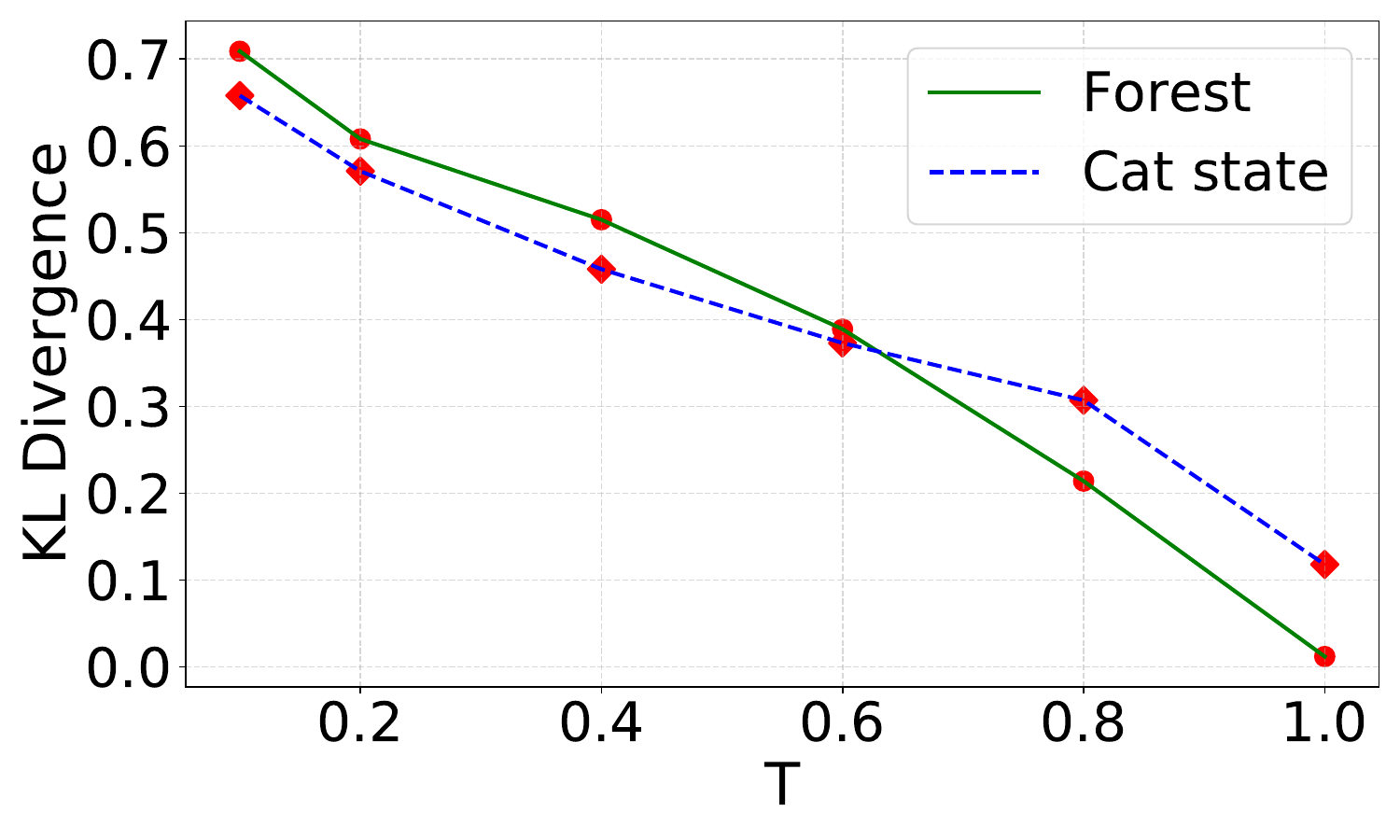}
    \caption{KL divergence (Eq.\ \eqref{eq:KL}) against the loss parameter ($T$) for the forest SAR image (classical) and cat state (quantum) data cases. Each data point corresponds to 5 runs.}
    \label{fig:KLplot}
\end{figure}

We analyzed the cost during training as a function of the energy transmissivity $T$ for one classical (forest SAR image as seen in Fig.\ \ref{fig:forest_f1}) and one quantum (non-Gaussian cat state \ Fig. \ref{fig:quantumdata}(b)) case. Results are shown in Fig.~\ref{fig:addingnoise}. Each loss parameter run was done 5 times. The mean of each run is displayed with a solid line with a shaded region around it representing the standard deviation. As expected, we obtained the best fidelity for forest (cat) 99.7\% (98.2\%) when $T=1$, corresponding to no loss. The fidelity decreases as we decrease the loss parameter, which makes sense because increasing the noise will make it harder for the machine to train. Eventually, at $T=0$, no training occurs. With every 10\% decrease in the loss, the fidelity changes by around 5\%. At 90\% loss in the channel, the fidelity change is around 30\%.  

In order to study how the loss channel impacts the performance of our CVQBM, we plotted the KL divergence \textit{vs.}\ the loss parameter ($T$) in Fig.~\ref{fig:KLplot}. In this plot, each data point corresponds to 5 training runs. We generated the distribution using the optimized parameters and hence calculated the KL divergence between the target and generated distributions (Eq.\ \eqref{eq:KL}). The lower value of KL divergence corresponds to the better match between the two distributions. Hence, the KL divergence decreases as $T\to 1$. As expected, best results are obtained at $T=1$ (no loss). The difference between the classical and quantum cases is small. We note that, up to 40\% loss, our CVQBM performs better the quantum case, but with more loss, the performance in the classical case is slightly better.

There is ongoing research for developing error mitigation strategies for photon loss in photonic hardware. For example, in Ref.\ \cite{mills2024mitigating}, techniques were presented that can mitigate the effects of photon loss on both output probabilities and expectation values in noisy linear optical circuits. These techniques involve the creation of recycled probabilities, which are derived from output statistics affected by loss and are designed to amplify the signal of the ideal probabilities. These recycled probabilities are then processed to produce a set of loss-mitigated probabilities or expectation values. These technique were shown to perform better than post-selection and zero-noise-extrapolation (ZNE) techniques. We leave study of photon loss mitigation strategies as a future work.

\section{Conclusion} \label{sec:conclude}
We proposed a CVQBM which can be realized experimentally with a CV photonic quantum computer. We introduced a novel method to prepare the thermal state required by the BM setup using an entangled state followed by QITE steps. We implemented the entire CVQBM design using Gaussian gates and photon number resolving measurements, making it experimentally realizable. 

We tested our CVQBM against both classical and quantum data. For classical data, we worked with the SAR images. These images are not affected by altitude, bad weather, or light, making them amenable to high-resolution remote sensing. We focused on datasets obtained from forest and sea ice monitoring. We generated distributions that were of very low KL divergence. All of the distributions we studied have various real-life applications ranging from SAR images \cite{gao2010statistical} to medical images \cite{slabaugh2009statistical}. For example, SAR images are obtained from  a wide-range of applications, such as agriculture monitoring, disaster management after damages done by a disaster like an earthquake, flood, and volcanic eruptions, search and rescue operations for the same, climate change studies, geology, and mining. 
For quantum data, we focused on a Gaussian distribution generated by a coherent state as well as one generated by a cat state, which have various applications, such as in designing qubits which are resistant to errors \cite{mirrahimi2014dynamically}, implementing entangling gates \cite{chen2022fault}, and quantum spectroscopy \cite{kira2011quantum}. 



In this work, we focused on generating continuous probability distributions that have many real-world applications. However, our proposed CVQBM can also be further utilized for anomaly detection \cite{bermot2023quantum} and classification tasks, as well as variational inference as discussed in \cite{benedetti2021variational} by optimizing a candidate probability distribution to approximate a posterior distribution over unobserved variables. This task has many real-world applications with continuous valued data that range from identifying diseases from symptoms to classifying economic regimes from price movements. Work in this direction is in progress.

\bmhead{Acknowledgements}

Research funded by the National Science Foundation under award DGE-2152168 and by the Department of Energy under award DE-SC0024325. 
A portion of the computation for this work was performed on the University of Tennessee Infrastructure for Scientific Applications and Advanced Computing (ISAAC) computational resources.

KYA was supported by MITRE's Quantum Horizon Program and by the Department of Energy under award DE-SC0024325. \copyright 2024 The MITRE Corporation. ALL RIGHTS RESERVED. Approved for public release. Distribution unlimited PR\textunderscore24$-$00320$-$3.




\end{document}